\documentclass[a4paper,11pt]{article}
\usepackage{pos}
\usepackage{lineno}
\usepackage{overpic}

\title{Joint searches by FACT, H.E.S.S., MAGIC and VERITAS for VHE gamma-ray emission associated with neutrinos detected by IceCube}
 \ShortTitle{Joint IACT observations of neutrinos detected by IceCube}

\author*[a]{Fabian Sch\"ussler}
\author[b]{H. Ashkar}
\author[c]{E. Bernardini}
\author[d]{A. Berti}
\author[a]{F. Bradascio}
\author[e]{S. Buson}
\author[e, f]{D. Dorner}
\author[g]{W. Jin}
\author[h]{G. Kukec Mezek}
\author[g]{M. Santander}
\author[i]{K. Satalecka}
\author[e, f]{B. Schleicher}
\author[h]{M. Senniappan}
\author[c]{I. Viale}

\affiliation[a]{IRFU, CEA, Université Paris-Saclay, Gif-sur-Yvette, France}

\affiliation[b]{Laboratoire Leprince-Ringuet, Ecole Polytechnique, CNRS, Institut Polytechnique de Paris, Palaiseau, France}
\affiliation[c]{Dipartimento di Fisica e Astronomia dell’Università and Sezione INFN, Padova, Italy, Padova, Italy}
\affiliation[d]{Max-Planck-Institut für Physik, München, München, Germany}
\affiliation[e]{Julius-Maximilians-Universität Würzburg, Fakultät für Physik und Astronomie, Institut für Theoretische Physik und Astrophysik, Lehrstuhl für Astronomie, Emil-Fischer-Str.\ 31, D-97074 Würzburg, Germany}
\affiliation[f]{ETH Zurich, Institute for Particle Physics and Astrophysics, Otto-Stern-Weg 5, 8093 Zürich, Switzerland}
\affiliation[g]{Department of Physics and Astronomy, University of Alabama, Tuscaloosa, Alabama, USA}
\affiliation[h]{Department of Physics and Electrical Engineering, Linnaeus University, Växjö, Sweden}
\affiliation[i]{Deutsches Elektronen-Synchrotron (DESY), Zeuthen, Germany}

\onbehalf{on behalf of the FACT, Fermi-LAT, H.E.S.S., IceCube, MAGIC and VERITAS collaborations} 

\emailAdd{fabian.schussler@cea.fr}

\abstract{The sources of the astrophysical flux of high-energy neutrinos detected by IceCube are still largely unknown, but searches for temporal and spatial correlation between neutrinos and electromagnetic radiation are a promising approach in this endeavor. All major imaging atmospheric Cherenkov telescopes (IACTs) - FACT, H.E.S.S., MAGIC, and VERITAS - operate an active follow-up program of target-of-opportunity observations of neutrino alerts issued by IceCube. These programs use several complementary neutrino alert streams. A publicly distributed alert stream is formed by individual high-energy neutrino candidate events of potentially astrophysical origin, such as IceCube-170922A (which could be linked to the flaring blazar TXS\,0506+056). A privately distributed alert stream is formed by clusters of neutrino events in time and space around either pre-selected gamma-ray sources or anywhere in the sky.
Here, we present joint searches for multi-wavelength emission associated with a set of IceCube alerts, both private and public, received through mid-January 2021. We will give an overview of the programs of the participating IACTs. We will showcase the various follow-up and data analysis strategies employed in response to the different alert types and various possible counterpart scenarios. Finally, we will present results from a combined analysis of the VHE gamma-ray observations obtained across all involved instruments, as well as relevant multi-wavelength data.}

\ConferenceLogo{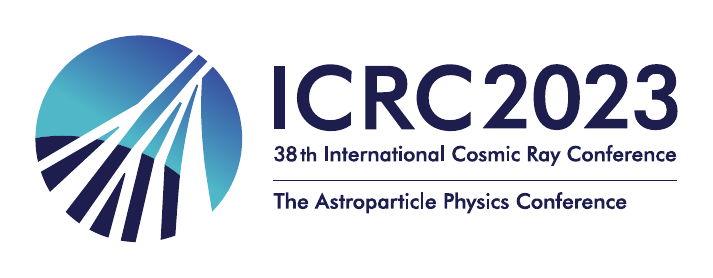}

\FullConference{%
38th International Cosmic Ray Conference (ICRC2023)\\
  26 July - 3 August, 2023\\
  Nagoya, Japan}

\begin{document}
\maketitle

\section{Introduction}

All currently operating imaging atmospheric Cherenkov telescopes (IACTs): FACT \citep{Biland:2014fqa}, H.E.S.S.\ \citep{2006A&A...457..899A}, MAGIC \citep{2012APh....35..435A} and VERITAS \citep{Holder:2006gi} operate observational programs in cooperation with IceCube. These target-of-opportunity (ToO) follow-up observations aim at identifying potential gamma-ray emission associated to the detected neutrino events. The programs can be broadly divided into two categories, depending on the type of IceCube neutrino alert that triggered the ToO observations. The first involves following up on individual high-energy ($>$60\,TeV) neutrino events that are likely of astrophysical origin, referred to as "singlets." These singlets are openly broadcasted by IceCube through AMON~\cite{AyalaSolares:2019jdz} via the NASA's General Coordinates Network (GCN)\footnote{\url{https://gcn.nasa.gov/}} on two streams, ``GOLD'' and ``BRONZE'', with an average astrophysical probability per event of 50\% and 30\%, respectively. IceCube has recently released a catalog of singlets of likely astrophysical origin up to December 2020~\cite{IceCube:2023agq}. The second type involves conducting follow-up observations of known gamma-ray sources for which IceCube has detected a cluster of candidate neutrino events, typically with energies around $\sim1$\,TeV, over a specific time period, known as "multiplets." These alerts are privately distributed by IceCube to individual IACTs under mutual agreements in the framework of the Gamma-ray Follow-Up (GFU) program~\citep{2016JInst..1111009I}.

Single-event alerts have a typical localization uncertainty of approximately $1^\circ$. This aligns well with the standard field of view (FOV) of imaging atmospheric Cherenkov telescopes (IACTs), which ranges from 3.5 to 5$^\circ$. Often, potential neutrino source candidates can be found within the region of interest (ROI) defined by the uncertainty of the neutrino localization. These candidates may include previously identified objects such as active galactic nuclei (AGN) like TXS 0506+056 from the $Fermi$-LAT and IACT catalogs, or transient sources detected by other electromagnetic (EM) observatories. In such cases, observations are typically focused on these objects. However, in the absence of pre-identified promising candidates, the searches generally aim to cover the entire ROI indicated by the neutrino localization uncertainty.

On the other hand, the primary objective of the GFU multiplet alerts is to notify the IACTs of neutrino flares from the direction of known gamma-ray emitters. The duration of these flares is not predetermined and can span from seconds to 180 days. Since the associated source of the alert is already known and identified as a GeV and/or TeV emitter, the IACT observations aim to determine possible changes to the state of the source (e.g.,\ quiescence vs. flaring or spectral changes). In its current configuration, a GFU multiplet alert is sent to the participating IACT if the statistical significance of the identified neutrino excess is greater than a predefined threshold, typically $3\sigma$ before accounting for statistical trials.

\begin{figure}[h!]

\begin{center}
\begin{overpic}[abs,unit=1mm, width= 0.75\textwidth]{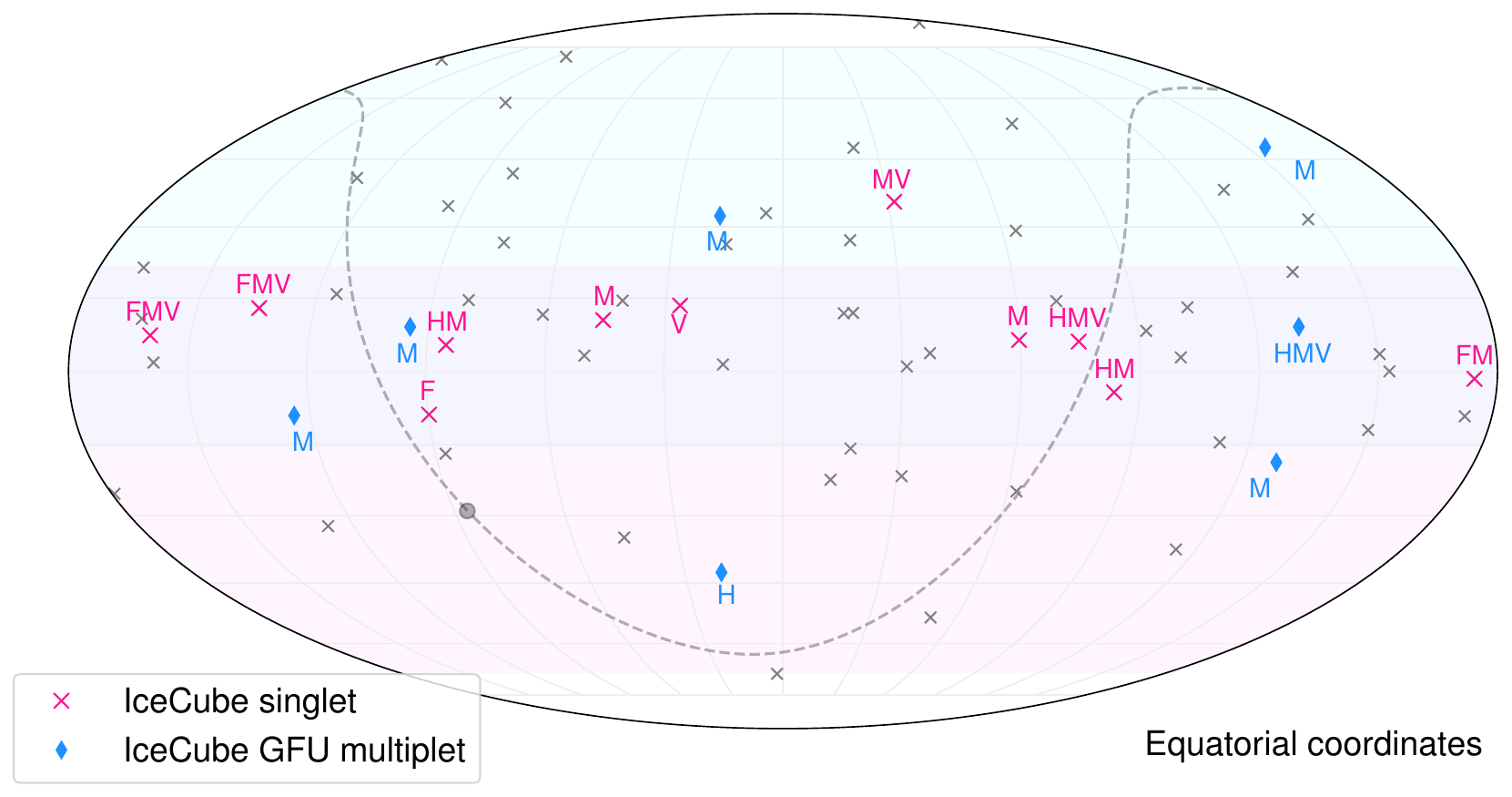}\put(3,55){\color{blue}\large preliminary} 
\end{overpic}
\caption{Sky map in equatorial coordinates showing IceCube alert positions observed by IACTs between October 2017 and January 2021 (in color, according to the alert type), and those not followed-up during the same period shown in gray. Letters indicate which IACTs participated in the observations (F - FACT, H - H.E.S.S., M - MAGIC, V - VERITAS). Light cyan and magenta bands indicate regions of the sky potentially observable at a zenith distance below 45$^{\circ}$ from the Northern (FACT, MAGIC, VERITAS) and Southern (H.E.S.S.) IACTs, respectively. The visibility windows for instruments in both hemispheres overlap around the celestial equator, where the IceCube sensitivity to neutrinos in the $\sim100$\,TeV-energy-range is maximal.}
\label{fig:map}
\end{center}
\end{figure}


The directions of the alerts sent by IceCube as neutrino event ``singlets'' and GFU neutrino {\it flares} between October 2017 and January 2021 are illustrated on a skymap in Figure~\ref{fig:map}. In the covered time period, the IACTs observed 11 of the 62 single-event alerts sent by IceCube. In the framework of the GFU program, IceCube sent 27 neutrino {\it flare} alerts from 17 sources, of which seven, including one ``all-sky alert", were observed. Here, we do not attempt to report a comprehensive summary of these alerts and follow-up observations, but rather will present a few examples and highlights to illustrate the various programs.

\section{Follow-up of neutrino multiplets: 1ES~1312-423}

The blazar known as 1ES~1312-423, located approximately 2 degrees away from the Centaurus~A (CenA) radio galaxy, has a redshift of $\mathrm{z} = 0.105 \pm 0.001$~\citep{2011NewA...16..503M}. Between April 2004 and July 2010, it has been observed by H.E.S.S. as part of the deep observation of Cen A, collecting a total exposure time of 150.6 h centered on CenA. This archival dataset revealed a power-law fit to the differential energy spectrum $\phi(E) = \mathrm{d}N/\mathrm{d}E$ of the VHE gamma-ray emission above $280~\mathrm{GeV}$. The best-fitting parameters obtained were $\Gamma = 2.85 \pm 0.47 (\mathrm{stat}) \pm 0.20 (\mathrm{sys})$ for the spectral index and a differential flux at 1\,TeV of $\phi(1~\mathrm{TeV}) = (1.89 \pm 0.58 (\mathrm{stat}) \pm 0.39 (\mathrm{sys})) \times 10^{-13}~\mathrm{cm}^{-2}~\mathrm{s}^{-1}~\mathrm{TeV}^{-1}$~\citep{2013MNRAS.434.1889H}.

\begin{figure}[th]
\begin{center}
\begin{overpic}[abs,unit=1mm, width= 0.6\textwidth]{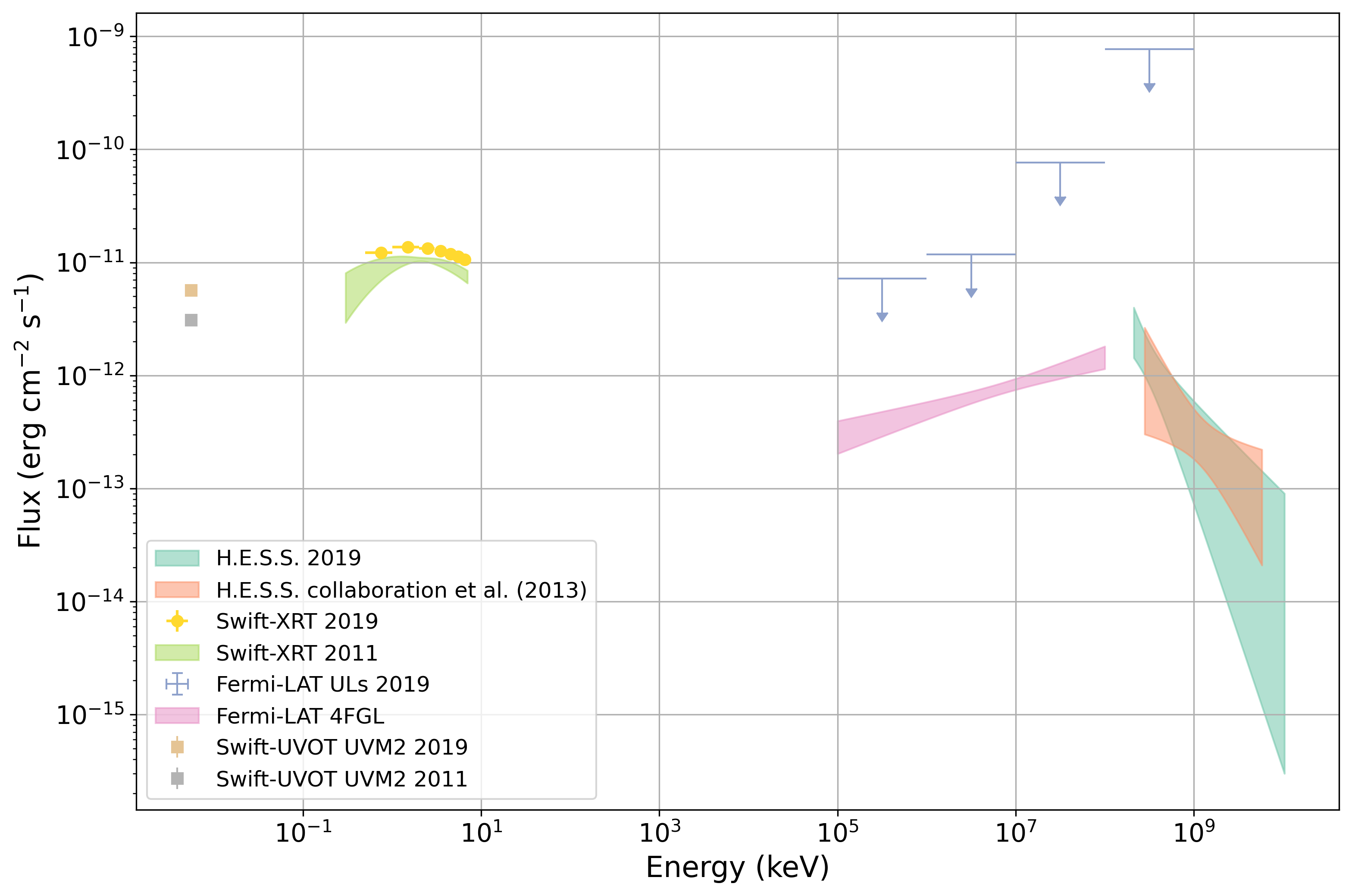}\put(10,55){\color{blue}\large preliminary} 
\end{overpic}
        \caption{1ES~1312-423 MWL SED showing archival data and observations obtained during the period following the GFU neutrino alert and contemporaneous to H.E.S.S. ToO observations.}\label{fig:1es1312}
    \end{center}
\end{figure}

Due to its Southern location 1ES~1312-423 is only part of the H.E.S.S.\ GFU program. On March 12, 2019, IceCube reported the detection of a short neutrino flare lasting $6.24\,\mathrm{h}$ from this source. H.E.S.S. conducted observations of the source for a total of $2.6~\mathrm{h}$ over two nights on March 12 and 13, 2019. 1ES~1312-423 was detected above $140~\mathrm{GeV}$ with a significance of $4\sigma$. This is the only detection of VHE gamma rays in any of the ToO campaigns conducted by the IACTs in the time period covered here. Again applying a power-law fit, the best-fit parameters $\Gamma = 2.73 \pm 0.27 (\mathrm{stat}) \pm 0.20 (\mathrm{sys})$ for the spectral index and $\phi(1~\mathrm{TeV}) = (5.83 \pm 3.2 (\mathrm{stat}) \pm 0.4 (\mathrm{sys})) \times 10^{-12}~\mathrm{cm}^{-2}~\mathrm{s}^{-1}~\mathrm{TeV}^{-1}$ for the differential flux at 1\,TeV were obtained. A comparison of the spectral energy distribution (SED) with the archival dataset is shown in Figure~\ref{fig:1es1312}, along with the results obtained from dedicated observations by {\it Swift} (UV + X-rays) and ATOM (optical). Although some variations are observed in the X-ray and UV domains, no clear change in the state of 1ES\,1312-423 could be identified in this neutrino-triggered ToO.

\section{Follow-up of neutrino singlets: IceCube-201114A + IceCube-171106A}

\begin{figure}
    \centering
    \begin{minipage}{0.47\textwidth}
        \centering
        \begin{overpic}[abs,unit=1mm, width= \textwidth]{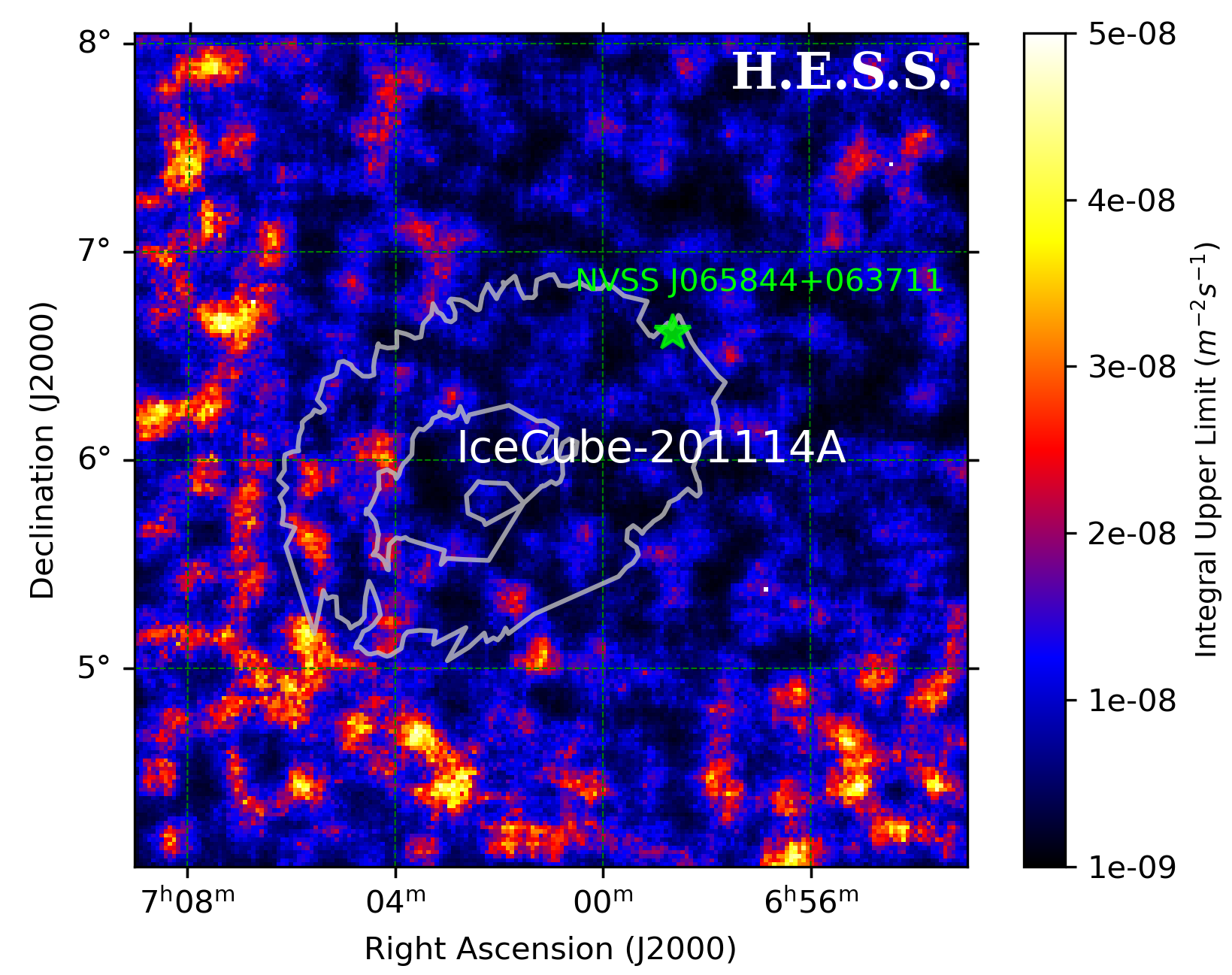}\put(9,8){\color{white}\normalsize preliminary} 
\end{overpic}
\caption{Integral upper limits on the VHE gamma-ray flux above 307\,GeV derived from the H.E.S.S.\ observations of IceCube-201114A. The observations were centered around the location of NVSS\,J065844+063711, indicated by the green star. The grey lines denote the 50\% (90\%) contours of the localization of the event by IceCube.}
\label{fig:201114A_HESSUL_contours}
    \end{minipage}\hfill
    \begin{minipage}{0.45\textwidth}
        \centering
        \begin{overpic}[abs,unit=1mm, width= \textwidth]{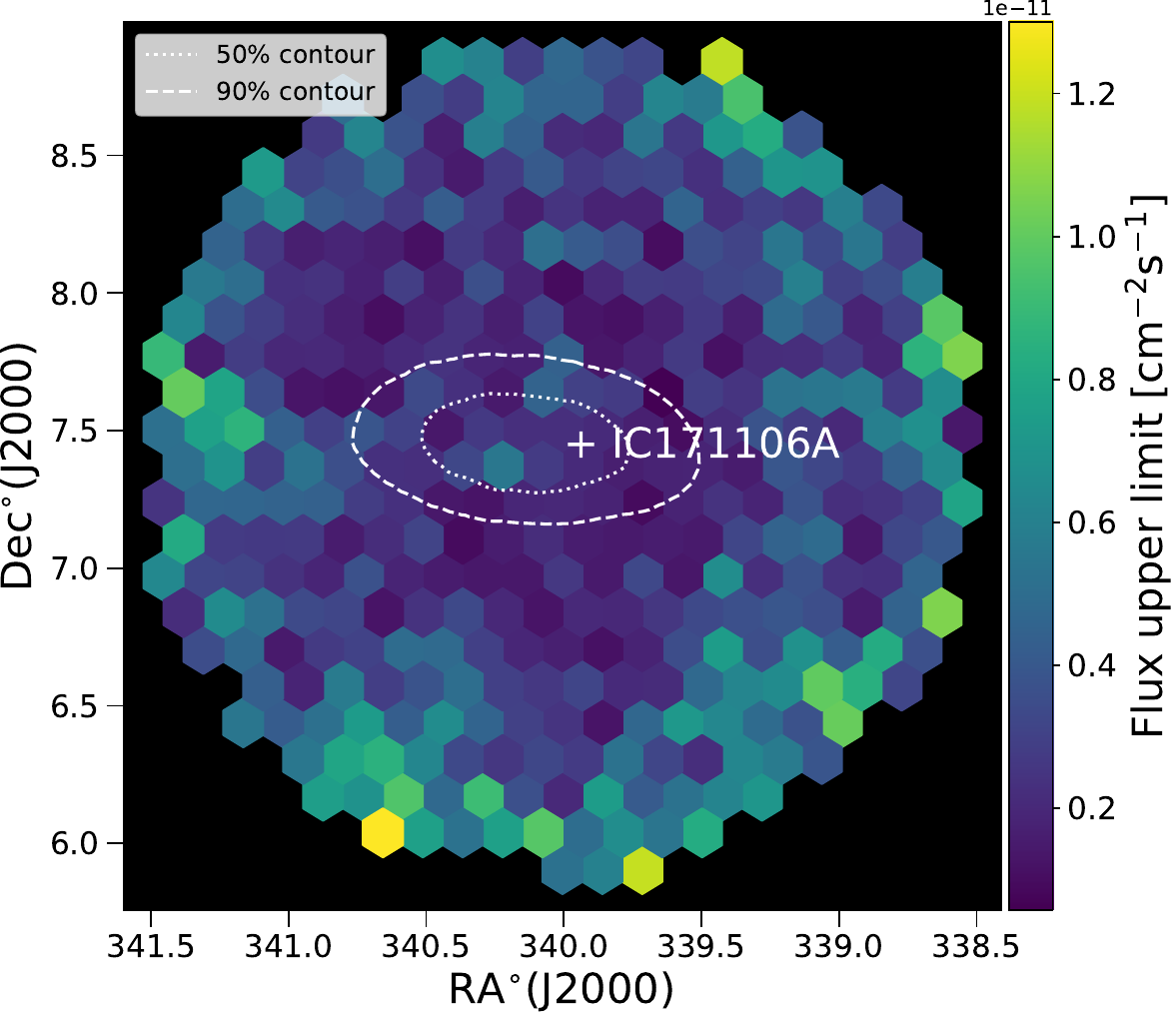}\put(9,8){\color{white}\normalsize preliminary} 
\end{overpic}
\caption{Map of integral upper limits of the VHE gamma-ray flux derived from the VERITAS observations of IceCube-171106A above an energy threshold of 350\,GeV. The dashed lines denote the 50\% (90\%) contours of the localization of the event by IceCube.}
\label{fig:IC171106A_VERITAS}
    \end{minipage}
\end{figure}

IceCube reported the detection of a track-like neutrino event classified as a GOLD alert on November 14th, 2020. The event had an energy of 214\,TeV and a probability to be of astrophysical origin of 0.56. Its false alert rate was determined to be 0.92 per year. The refined direction of the alert was determined to be RA: 105.25$^\circ$ (+1.28$^\circ$, $-$1.12$^\circ$, 90\% PSF containment) and Dec: 6.05$^\circ$ (+0.95$^\circ$ $-$0.95$^\circ$, 90\% PSF containment). The position of the neutrino event is compatible with the blazar NVSS\,J065844+063711, also known as the Fermi source 4FGL\,J0658.6+0636 (RA: 104.64$^\circ$, Dec: 6.60$^\circ$). A map of the region is shown in Figure~\ref{fig:201114A_HESSUL_contours}.

The event was observed by H.E.S.S., MAGIC, and VERITAS. H.E.S.S. collected 14.3 hours of data during an observation campaign conducted from November 18th to November 25th, 2020, and from December 10th, 2020, to December 11th, 2020. No significant emission was detected in the ROI, and integral upper limits on the VHE gamma-ray flux were derived at the position of NVSS\,J065844+063711 as $5.37\times10^{-13}\, \mathrm{cm^{-2}} \, \mathrm{s^{-1}}$ above 326\,GeV. The MAGIC telescopes observed the direction of NVSS\,J065844+063711 on November 16th, 17th, and from November 19th to 25th, 2020. The 6\,h of data did not result in a detection. The integral flux upper limit was calculated above an energy threshold of 120\,GeV to be $8.39\times10^{-10}\,\mathrm{cm^{-2}} \, \mathrm{s^{-1}}$. VERITAS conducted observations from November 15th to November 19th, 2020, collecting approximately 7 hours of quality-selected observation data. No detection was made at the location indicated by the IceCube neutrino event or elsewhere in the VERITAS field of view. The integral flux upper limit above an energy threshold of 200\,GeV at the position of the blazar NVSS\,J065844+063711 was determined to be $1.37\times10^{-12}\, \mathrm{cm^{-2}} \, \mathrm{s^{-1}}$.

\begin{figure}
    \centering
    \begin{minipage}{0.45\textwidth}
        \centering
        \begin{overpic}[abs,unit=1mm, width= 0.9\textwidth]{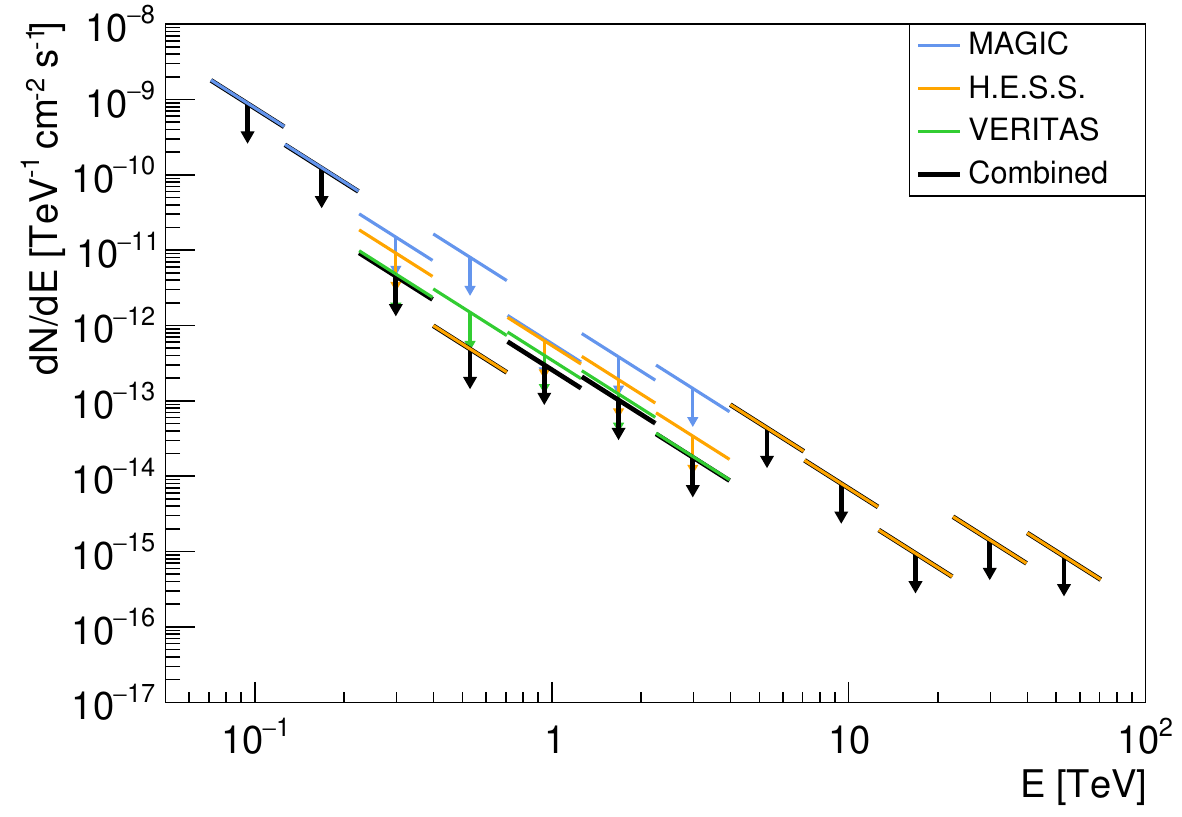}\put(11,8){\color{blue}\normalsize preliminary} 
\end{overpic}
        \caption{Combined differential flux upper limits on the VHE gamma-ray flux (black arrows) obtained from observations of IceCube-201114A by multiple IACTs (colored arrows).}\label{fig:ul_combined}
    \end{minipage}\hfill
    \begin{minipage}{0.45\textwidth}
        \centering
        \begin{overpic}[abs,unit=1mm, width= \textwidth]{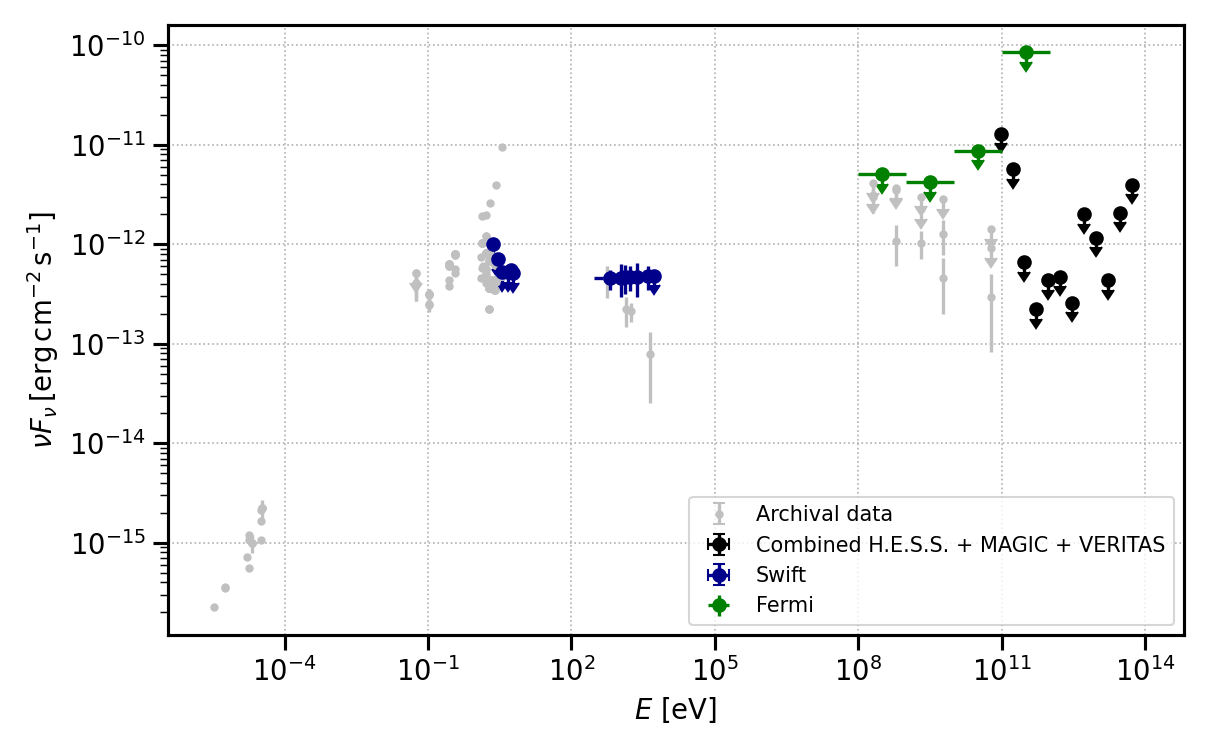}\put(11,8){\color{blue}\normalsize preliminary} 
\end{overpic}
        \caption{Multi-wavelength spectal energy distribution for NVSS\,J065844+063711 (4FGL\,J0658.6+0636) within the localisation uncertainty of IceCube-201114A. }\label{fig:mwl}
    \end{minipage}
\end{figure}

The profile maximum likelihood approach described in~\cite{MAXIJ1820} is employed to generate combined flux upper limits across all IACTs. For this purpose, all collaborations utilized the same energy binning, dividing the VHE range between 71\,GeV and 71\,TeV into four bins per decade. We used the Rolke method~\citep{Lundberg_2010} with a confidence level set to 95\%, and including a 30\% global systematic uncertainty in the efficiency of the applied cuts. Upper limits were calculated considering an observed spectrum modeled by a power law d$N/$d$E = K E^{-\Gamma}$ with index $\Gamma = 2.5$. The individual as well the combined upper limits at the position of of the blazar NVSS\,J065844+063711 are shown in Figure~\ref{fig:ul_combined}.

Complementing the VHE gamma-ray observations, {\it Swift}-XRT observations were triggered on November 15th, 2020. They revealed a slight increase in the X-ray flux around the time of the neutrino emission compared to previous observations in May 2012. The X-ray flux level remained relatively consistent until December 11th, 2020. A full MWL SED is shown in Figure~\ref{fig:mwl}.

Single-event public IceCube alerts are characterized by relatively large localization errors ($\sim$\,0.5 degree to a few degrees at 50\% confidence level). As a consequence, the exact localization of the neutrino origin can not be pin-pointed to any point-source with high accuracy. Therefore, we derive sky maps containing the integral gamma-ray flux upper limits and covering the neutrino arrival localization uncertainty. Dedicated searches for a variety of possible source candidates in this region are thus possible. Examples for these upper limit maps are given in Figure~\ref{fig:201114A_HESSUL_contours} and Figure~\ref{fig:IC171106A_VERITAS}. The latter has been derived above an energy threshold of 350\,GeV from about 2.5~hours of quality-selected observation data taken by VERITAS between November 13 to November 20, 2017. This ToO, that was also followed by MAGIC and FACT, was triggered following the detection of IceCube-171106A on November 6th, 2017, an event with an energy of 230\,TeV and a signalness indicating the probability of the event to be of astrophysical origin of 75\%.

\section{Conclusions}
\begin{figure}[th]
\begin{center}
\begin{overpic}[abs,unit=1mm, width=0.8\textwidth]{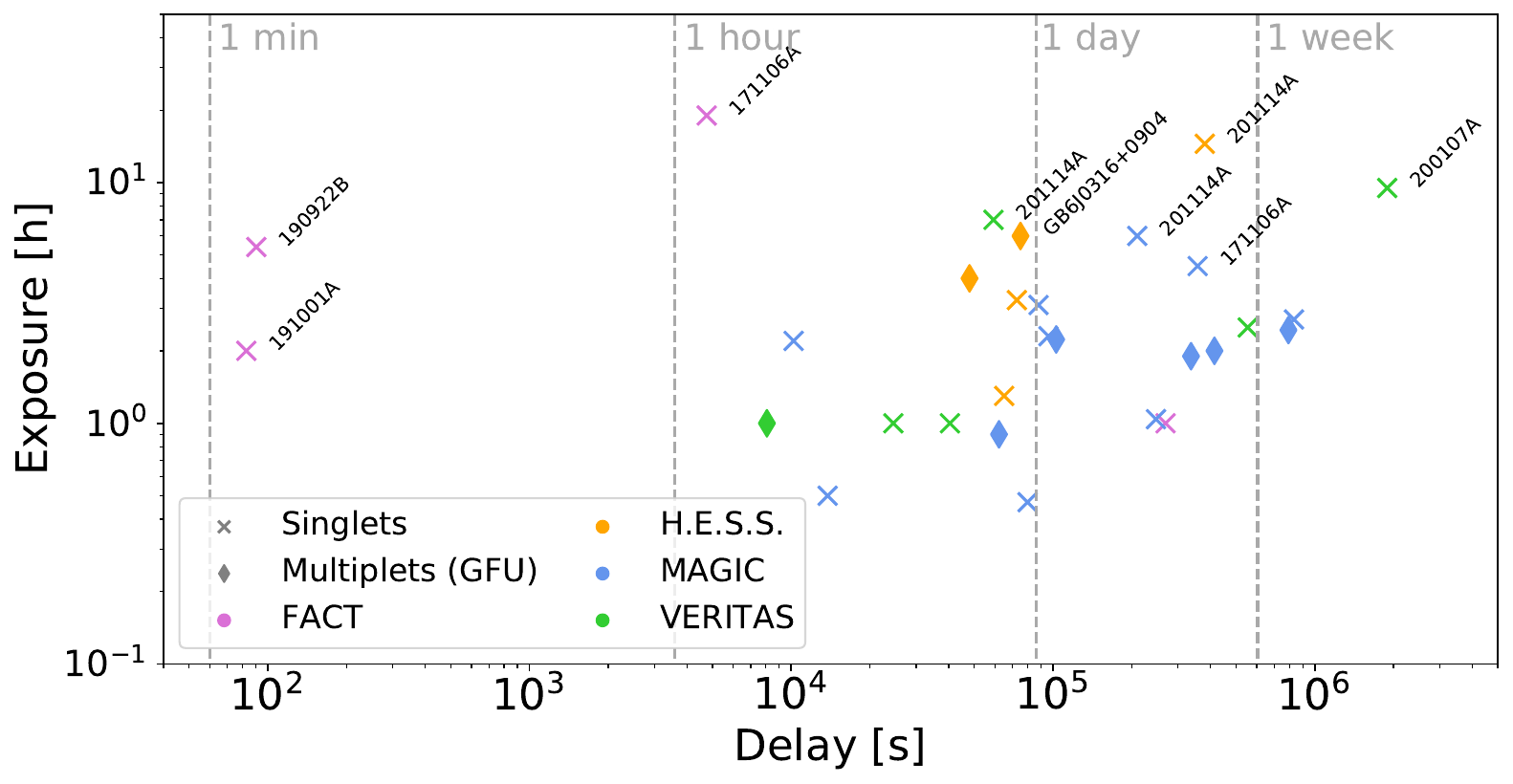}\put(98,12){\color{blue}\large preliminary} 
\end{overpic}
\caption{Delay vs exposure times for IACT follow-up of neutrino alerts from October 2017 until December 2020. The delay is calculated from the neutrino event arrival time (single events) or flare threshold crossing time (multiplets) up to the start of the IACT observation. Highlighted are observations performed with a start delay less than 100\,s or with a total exposure longer than 4\,h. Marker color represents the IACT observing while the marker type represents the alert type.}
\label{fig:delay}
\end{center}
\end{figure}

Figure~\ref{fig:delay} shows the delay and exposure of the IACT observations of single-event alerts and GFU neutrino flares between October 2017 and January 2021. The delay is derived from the detection time of the neutrino event (for single events) or the threshold crossing time of the flare (for multiplets) up to the start of IACT observation.

From this overview, several general trends in the follow-up strategies can be inferred. Approximately 50\% of the cases achieve a reaction within one day, while observations conducted more than one week after the trigger are rare. The total time allocated to the follow-up of public IceCube alerts is comparable across all collaborations, amounting to approximately 20 hours for the covered period. However, there are differences in the approaches taken. H.E.S.S., and VERITAS primarily focus on longer exposures for a few selected alerts, whereas MAGIC conducts a larger number of follow-ups with shorter average exposure. Similar trends are observed for GFU neutrino alerts.

The geographic distribution of the IACTs in terms of latitude has facilitated complete coverage of the entire sky, encompassing both the Northern and Southern hemispheres. Moreover, the observatories' locations in longitude have expanded the field of regard of the combined network of IACTs, thereby increasing the likelihood of promptly conducting follow-up observations. This is particularly crucial in cases where observability from a single observatory site is restricted due to factors such as adverse weather conditions, the presence of the sun or moon, or technical issues. This aspect holds significant importance, as it enables the observation of VHE gamma rays associated with rare transient neutrino candidate events. It also allows for the exploration of other time-domain or multi-messenger triggers, as exemplified by follow-up observations of gravitational wave events (e.g., GW170817A) or gamma-ray bursts (e.g., VHE-detected GRBs and GRB 221009A). These examples underscore the value of conducting joint analyses, as presented here, to fully leverage the capabilities and strengths of the collective observatory network.


\bibliography{IACT-NuToO_ICRC2023}{}
\bibliographystyle{unsrt} %


%
%
%

We acknowledge support by Institut Pascal at Université Paris-Saclay during the Paris-Saclay Astroparticle Symposium 2021 and 2022, with the support of the P2IO Laboratory of Excellence (program “Investissements d’avenir” ANR-11-IDEX-0003-01 Paris-Saclay and ANR-10-LABX-0038), the P2I axis of the Graduate School Physics of Université Paris-Saclay, as well as IJCLab, CEA, IPhT, IAS, OSUPS, the IN2P3 master projet UCMN, APPEC, and EuCAPT.

F. Sch\"ussler and H. Ashkar acknowledge the support of the French Agence Nationale de la Recherche (ANR) under reference ANR-22-CE31-0012. This work was also supported by the Programme National des Hautes Energies of CNRS/INSU with INP and IN2P3, co-funded by CEA and CNES.

This work was supported by the European Research Council, ERC Starting grant MessMapp, S.Buson Principal Investigator, under contract no. 949555.

\section*{Full Author List: FACT Collaboration}

\scriptsize
\noindent
J.~Abhir $^{1}$, 
D.~Baack$^{2}$,
M.~Balbo $^{3}$, 
A.~Biland $^{1}$, 
K.~Brand $^{4}$, 
T.~Bretz $^{1}$ $^{a}$, 
J.~Buss $^{2}$, 
D.~Dorner $^{1,4}$, 
L.~Eisenberger $^{4}$, 
D.~Elsaesser $^{2}$, 
P.~Günther $^{4}$, 
D.~Hildebrand $^{1}$,
K.~Mannheim $^{4}$, 
M.~Noethe $^{2}$, 
A.~Paravac $^{4}$, 
W.~Rhode $^{2}$, 
B.~Schleicher $^{1,4}$, 
V.~Sliusar $^{3}$, 
S.~Hasan $^{1}$,
R.~Walter $^{3}$
\\

\noindent
$^1$ ETH Zurich, Institute for Particle Physics and Astrophysics, Otto-Stern-Weg 5, 8093 Zürich, Switzerland \\
$^2$ TU Dortmund, Experimental Physics 5, Otto-Hahn-Str.\ 4a, 44227 Dortmund, Germany \\
$^3$ University of Geneva, Department of Astronomy, Chemin d'Ecogia 16, 1290 Versoix, Switzerland \\
$^4$ Julius-Maximilians-Universität Würzburg, Fakultät für Physik und Astronomie, Institut für Theoretische Physik und Astrophysik, Lehrstuhl für Astronomie, Emil-Fischer-Str.\ 31, D-97074 Würzburg, Germany\\
$^a$ also at RWTH Aachen University \\ 

\subsection*{Acknowledgements}

\noindent
 The important contributions from ETH Zurich grants ETH-10.08-2 and ETH-27.12-1 as well as the funding by the Swiss SNF and the German BMBF (Verbundforschung Astro- und Astroteilchenphysik) and HAP (Helmoltz Alliance for Astroparticle Physics) are gratefully acknowledged. Part of this work is supported by Deutsche Forschungsgemeinschaft (DFG) within the Collaborative Research Center SFB 876 "Providing Information by Resource-Constrained Analysis", project C3. We are thankful for the very valuable contributions from E. Lorenz, D. Renker and G. Viertel during the early phase of the project. We thank the Instituto de Astrofísica de Canarias for allowing us to operate the telescope at the Observatorio del Roque de los Muchachos in La Palma, the Max-Planck-Institut für Physik for providing us with the mount of the former HEGRA CT3 telescope, and the MAGIC collaboration for their support. 
\section*{Acknowledgements: \textit{Fermi} LAT Collaboration}

The \textit{Fermi} LAT Collaboration acknowledges generous ongoing support
from a number of agencies and institutes that have supported both the
development and the operation of the LAT as well as scientific data analysis.
These include the National Aeronautics and Space Administration and the
Department of Energy in the United States, the Commissariat \`a l'Energie Atomique
and the Centre National de la Recherche Scientifique / Institut National de Physique
Nucl\'eaire et de Physique des Particules in France, the Agenzia Spaziale Italiana
and the Istituto Nazionale di Fisica Nucleare in Italy, the Ministry of Education,
Culture, Sports, Science and Technology (MEXT), High Energy Accelerator Research
Organization (KEK) and Japan Aerospace Exploration Agency (JAXA) in Japan, and
the K.~A.~Wallenberg Foundation, the Swedish Research Council and the
Swedish National Space Board in Sweden.
 
Additional support for science analysis during the operations phase is gratefully
acknowledged from the Istituto Nazionale di Astrofisica in Italy and the Centre
National d'\'Etudes Spatiales in France. This work performed in part under DOE
Contract DE-AC02-76SF00515.
\section*{Full Author List: H.E.S.S. Collaboration}
\scriptsize
\noindent
F.~Aharonian$^{1,2,3}$, 
F.~Ait~Benkhali$^{4}$, 
A.~Alkan$^{5}$, 
J.~Aschersleben$^{6}$, 
H.~Ashkar$^{7}$, 
M.~Backes$^{8,9}$, 
A.~Baktash$^{10}$, 
V.~Barbosa~Martins$^{11}$, 
A.~Barnacka$^{12}$, 
J.~Barnard$^{13}$
% ...

\medskip

\noindent
$^{1}$Dublin Institute for Advanced Studies, 31 Fitzwilliam Place, Dublin 2, Ireland\\
$^{2}$Max-Planck-Institut f\"ur Kernphysik, P.O. Box 103980, D 69029 Heidelberg, Germany\\
$^{3}$Yerevan State University,  1 Alek Manukyan St, Yerevan 0025, Armenia\\
$^{4}$Landessternwarte, Universit\"at Heidelberg, K\"onigstuhl, D 69117 Heidelberg, Germany\\
$^{5}$IRFU, CEA, Universit\'e Paris-Saclay, F-91191 Gif-sur-Yvette, France\\
$^{6}$Kapteyn Astronomical Institute, University of Groningen, Landleven 12, 9747 AD Groningen, The Netherlands\\
$^{7}$Laboratoire Leprince-Ringuet, École Polytechnique, CNRS, Institut Polytechnique de Paris, F-91128 Palaiseau, France\\
$^{8}$University of Namibia, Department of Physics, Private Bag 13301, Windhoek 10005, Namibia\\
$^{9}$Centre for Space Research, North-West University, Potchefstroom 2520, South Africa\\
$^{10}$Universit\"at Hamburg, Institut f\"ur Experimentalphysik, Luruper Chaussee 149, D 22761 Hamburg, Germany\\
$^{11}$DESY, D-15738 Zeuthen, Germany\\
$^{12}$Obserwatorium Astronomiczne, Uniwersytet Jagiello{\'n}ski, ul. Orla 171, 30-244 Krak{\'o}w, Poland\\
$^{13}$Department of Physics, University of the Free State,  PO Box 339, Bloemfontein 9300, South Africa\\
% ...

\subsection*{Acknowledgements}

\noindent
The support of the Namibian authorities and of the University of
Namibia in facilitating the construction and operation of H.E.S.S.
is gratefully acknowledged, as is the support by the German
Ministry for Education and Research (BMBF), the Max Planck Society,
the Helmholtz Association, the French Ministry of
Higher Education, Research and Innovation, the Centre National de
la Recherche Scientifique (CNRS/IN2P3 and CNRS/INSU), the
Commissariat à l’énergie atomique et aux énergies alternatives
(CEA), the U.K. Science and Technology Facilities Council (STFC),
the Irish Research Council (IRC) and the Science Foundation Ireland
(SFI), the Polish Ministry of Education and Science, agreement no.
2021/WK/06, the South African Department of Science and Innovation and
National Research Foundation, the University of Namibia, the National
Commission on Research, Science \& Technology of Namibia (NCRST),
the Austrian Federal Ministry of Education, Science and Research
and the Austrian Science Fund (FWF), the Australian Research
Council (ARC), the Japan Society for the Promotion of Science, the
University of Amsterdam and the Science Committee of Armenia grant
21AG-1C085. We appreciate the excellent work of the technical
support staff in Berlin, Zeuthen, Heidelberg, Palaiseau, Paris,
Saclay, Tübingen and in Namibia in the construction and operation
of the equipment. This work benefited from services provided by the
H.E.S.S. Virtual Organisation, supported by the national resource
providers of the EGI Federation.
\section*{Full Author List: IceCube Collaboration}

\scriptsize
\noindent
R. Abbasi$^{17}$,
M. Ackermann$^{63}$,
J. Adams$^{18}$,
S. K. Agarwalla$^{40,\: 64}$,
J. A. Aguilar$^{12}$,
M. Ahlers$^{22}$,
J.M. Alameddine$^{23}$,
N. M. Amin$^{44}$,
K. Andeen$^{42}$,
G. Anton$^{26}$
% ...
\\
\\
% ...
$^{12}$ Universit{\'e} Libre de Bruxelles, Science Faculty CP230, B-1050 Brussels, Belgium \\
% ...
$^{17}$ Department of Physics, Loyola University Chicago, Chicago, IL 60660, USA \\
$^{18}$ Dept. of Physics and Astronomy, University of Canterbury, Private Bag 4800, Christchurch, New Zealand \\
% ...
$^{22}$ Niels Bohr Institute, University of Copenhagen, DK-2100 Copenhagen, Denmark \\
$^{23}$ Dept. of Physics, TU Dortmund University, D-44221 Dortmund, Germany \\
% ...
$^{26}$ Erlangen Centre for Astroparticle Physics, Friedrich-Alexander-Universit{\"a}t Erlangen-N{\"u}rnberg, D-91058 Erlangen, Germany \\
% ...
$^{40}$ Dept. of Physics and Wisconsin IceCube Particle Astrophysics Center, University of Wisconsin{\textendash}Madison, Madison, WI 53706, USA \\
% ...
$^{42}$ Department of Physics, Marquette University, Milwaukee, WI, 53201, USA \\
% ...
$^{44}$ Bartol Research Institute and Dept. of Physics and Astronomy, University of Delaware, Newark, DE 19716, USA \\
% ...
$^{63}$ Deutsches Elektronen-Synchrotron DESY, Platanenallee 6, 15738 Zeuthen, Germany  \\
$^{64}$ Institute of Physics, Sachivalaya Marg, Sainik School Post, Bhubaneswar 751005, India \\
% ...

\subsection*{Acknowledgements}

\noindent
The authors gratefully acknowledge the support from the following agencies and institutions:
USA {\textendash} U.S. National Science Foundation-Office of Polar Programs,
U.S. National Science Foundation-Physics Division,
U.S. National Science Foundation-EPSCoR,
Wisconsin Alumni Research Foundation,
Center for High Throughput Computing (CHTC) at the University of Wisconsin{\textendash}Madison,
Open Science Grid (OSG),
Advanced Cyberinfrastructure Coordination Ecosystem: Services {\&} Support (ACCESS),
Frontera computing project at the Texas Advanced Computing Center,
U.S. Department of Energy-National Energy Research Scientific Computing Center,
Particle astrophysics research computing center at the University of Maryland,
Institute for Cyber-Enabled Research at Michigan State University,
and Astroparticle physics computational facility at Marquette University;
Belgium {\textendash} Funds for Scientific Research (FRS-FNRS and FWO),
FWO Odysseus and Big Science programmes,
and Belgian Federal Science Policy Office (Belspo);
Germany {\textendash} Bundesministerium f{\"u}r Bildung und Forschung (BMBF),
Deutsche Forschungsgemeinschaft (DFG),
Helmholtz Alliance for Astroparticle Physics (HAP),
Initiative and Networking Fund of the Helmholtz Association,
Deutsches Elektronen Synchrotron (DESY),
and High Performance Computing cluster of the RWTH Aachen;
Sweden {\textendash} Swedish Research Council,
Swedish Polar Research Secretariat,
Swedish National Infrastructure for Computing (SNIC),
and Knut and Alice Wallenberg Foundation;
European Union {\textendash} EGI Advanced Computing for research;
Australia {\textendash} Australian Research Council;
Canada {\textendash} Natural Sciences and Engineering Research Council of Canada,
Calcul Qu{\'e}bec, Compute Ontario, Canada Foundation for Innovation, WestGrid, and Compute Canada;
Denmark {\textendash} Villum Fonden, Carlsberg Foundation, and European Commission;
New Zealand {\textendash} Marsden Fund;
Japan {\textendash} Japan Society for Promotion of Science (JSPS)
and Institute for Global Prominent Research (IGPR) of Chiba University;
Korea {\textendash} National Research Foundation of Korea (NRF);
Switzerland {\textendash} Swiss National Science Foundation (SNSF);
United Kingdom {\textendash} Department of Physics, University of Oxford.
\clearpage
\section*{Full Authors List: MAGIC Collaboration}
\scriptsize
\noindent
H.~Abe$^{1}$,
S.~Abe$^{1}$,
J.~Abhir$^{2}$,
V.~A.~Acciari$^{3}$,
I.~Agudo$^{4}$,
T.~Aniello$^{5}$,
S.~Ansoldi$^{6,46}$,
L.~A.~Antonelli$^{5}$,
A.~Arbet Engels$^{7}$,
C.~Arcaro$^{8}$
% ...
\noindent
$^{1}$ {Japanese MAGIC Group: Institute for Cosmic Ray Research (ICRR), The University of Tokyo, Kashiwa, 277-8582 Chiba, Japan} \\
$^{2}$ {ETH Z\"urich, CH-8093 Z\"urich, Switzerland} \\
$^{3}$ {Instituto de Astrof\'isica de Canarias and Dpto. de  Astrof\'isica, Universidad de La Laguna, E-38200, La Laguna, Tenerife, Spain} \\
$^{4}$ {Instituto de Astrof\'isica de Andaluc\'ia-CSIC, Glorieta de la Astronom\'ia s/n, 18008, Granada, Spain} \\
$^{5}$ {National Institute for Astrophysics (INAF), I-00136 Rome, Italy} \\
$^{6}$ {Universit\`a di Udine and INFN Trieste, I-33100 Udine, Italy} \\
$^{7}$ {Max-Planck-Institut f\"ur Physik, D-80805 M\"unchen, Germany} \\
$^{8}$ {Universit\`a di Padova and INFN, I-35131 Padova, Italy} \\
% ...
$^{46}$ {also at International Center for Relativistic Astrophysics (ICRA), Rome, Italy} \\
% ...

\section*{Full Author List: VERITAS Collaboration}

\scriptsize
\noindent
A.~Acharyya$^{1}$,
C.~B.~Adams$^{2}$,
A.~Archer$^{3}$,
P.~Bangale$^{4}$,
J.~T.~Bartkoske$^{5}$,
P.~Batista$^{6}$,
W.~Benbow$^{7}$,
J.~L.~Christiansen$^{8}$,
A.~J.~Chromey$^{7}$,
A.~Duerr$^{5}$,
M.~Errando$^{9}$,
Q.~Feng$^{7}$,
G.~M.~Foote$^{4}$,
L.~Fortson$^{10}$,
A.~Furniss$^{11, 12}$,
W.~Hanlon$^{7}$,
O.~Hervet$^{12}$,
C.~E.~Hinrichs$^{7,13}$,
J.~Hoang$^{12}$,
J.~Holder$^{4}$,
Z.~Hughes$^{9}$,
T.~B.~Humensky$^{14,15}$,
W.~Jin$^{1}$,
M.~N.~Johnson$^{12}$,
M.~Kertzman$^{3}$,
M.~Kherlakian$^{6}$,
D.~Kieda$^{5}$,
T.~K.~Kleiner$^{6}$,
N.~Korzoun$^{4}$,
S.~Kumar$^{14}$,
M.~J.~Lang$^{16}$,
M.~Lundy$^{17}$,
G.~Maier$^{6}$,
C.~E~McGrath$^{18}$,
M.~J.~Millard$^{19}$,
C.~L.~Mooney$^{4}$,
P.~Moriarty$^{16}$,
R.~Mukherjee$^{20}$,
S.~O'Brien$^{17,21}$,
R.~A.~Ong$^{22}$,
N.~Park$^{23}$,
C.~Poggemann$^{8}$,
M.~Pohl$^{24,6}$,
E.~Pueschel$^{6}$,
J.~Quinn$^{18}$,
P.~L.~Rabinowitz$^{9}$,
K.~Ragan$^{17}$,
P.~T.~Reynolds$^{25}$,
D.~Ribeiro$^{10}$,
E.~Roache$^{7}$,
J.~L.~Ryan$^{22}$,
I.~Sadeh$^{6}$,
L.~Saha$^{7}$,
M.~Santander$^{1}$,
G.~H.~Sembroski$^{26}$,
R.~Shang$^{20}$,
M.~Splettstoesser$^{12}$,
A.~K.~Talluri$^{10}$,
J.~V.~Tucci$^{27}$,
V.~V.~Vassiliev$^{22}$,
A.~Weinstein$^{28}$,
D.~A.~Williams$^{12}$,
S.~L.~Wong$^{17}$,
and
J.~Woo$^{29}$\\
\\
\noindent
$^{1}${Department of Physics and Astronomy, University of Alabama, Tuscaloosa, AL 35487, USA}

\noindent
$^{2}${Physics Department, Columbia University, New York, NY 10027, USA}

\noindent
$^{3}${Department of Physics and Astronomy, DePauw University, Greencastle, IN 46135-0037, USA}

\noindent
$^{4}${Department of Physics and Astronomy and the Bartol Research Institute, University of Delaware, Newark, DE 19716, USA}

\noindent
$^{5}${Department of Physics and Astronomy, University of Utah, Salt Lake City, UT 84112, USA}

\noindent
$^{6}${DESY, Platanenallee 6, 15738 Zeuthen, Germany}

\noindent
$^{7}${Center for Astrophysics $|$ Harvard \& Smithsonian, Cambridge, MA 02138, USA}

\noindent
$^{8}${Physics Department, California Polytechnic State University, San Luis Obispo, CA 94307, USA}

\noindent
$^{9}${Department of Physics, Washington University, St. Louis, MO 63130, USA}

\noindent
$^{10}${School of Physics and Astronomy, University of Minnesota, Minneapolis, MN 55455, USA}

\noindent
$^{11}${Department of Physics, California State University - East Bay, Hayward, CA 94542, USA}

\noindent
$^{12}${Santa Cruz Institute for Particle Physics and Department of Physics, University of California, Santa Cruz, CA 95064, USA}

\noindent
$^{13}${Department of Physics and Astronomy, Dartmouth College, 6127 Wilder Laboratory, Hanover, NH 03755 USA}

\noindent
$^{14}${Department of Physics, University of Maryland, College Park, MD, USA }

\noindent
$^{15}${NASA GSFC, Greenbelt, MD 20771, USA}

\noindent
$^{16}${School of Natural Sciences, University of Galway, University Road, Galway, H91 TK33, Ireland}

\noindent
$^{17}${Physics Department, McGill University, Montreal, QC H3A 2T8, Canada}

\noindent
$^{18}${School of Physics, University College Dublin, Belfield, Dublin 4, Ireland}

\noindent
$^{19}${Department of Physics and Astronomy, University of Iowa, Van Allen Hall, Iowa City, IA 52242, USA}

\noindent
$^{20}${Department of Physics and Astronomy, Barnard College, Columbia University, NY 10027, USA}

\noindent
$^{21}${ Arthur B. McDonald Canadian Astroparticle Physics Research Institute, 64 Bader Lane, Queen's University, Kingston, ON Canada, K7L 3N6}

\noindent
$^{22}${Department of Physics and Astronomy, University of California, Los Angeles, CA 90095, USA}

\noindent
$^{23}${Department of Physics, Engineering Physics and Astronomy, Queen's University, Kingston, ON K7L 3N6, Canada}

\noindent
$^{24}${Institute of Physics and Astronomy, University of Potsdam, 14476 Potsdam-Golm, Germany}

\noindent
$^{25}${Department of Physical Sciences, Munster Technological University, Bishopstown, Cork, T12 P928, Ireland}

\noindent
$^{26}${Department of Physics and Astronomy, Purdue University, West Lafayette, IN 47907, USA}

\noindent
$^{27}${Department of Physics, Indiana University-Purdue University Indianapolis, Indianapolis, IN 46202, USA}

\noindent
$^{28}${Department of Physics and Astronomy, Iowa State University, Ames, IA 50011, USA}

\noindent
$^{29}${Columbia Astrophysics Laboratory, Columbia University, New York, NY 10027, USA}

\end{document}